\documentclass[preprint]{epl}
\newcommand{\journal}[4]{{#1~}{#2}\,(#3)\,#4.}

\newcommand{\pr}{\journal {Phys. Rev.}}

\newcommand{\PL}{\journal {Phys. Lett.}}

\newcommand{\prep}{\journal {Phys. Reports}}

\newcommand{\PR}{\journal{Phys. Rev.}}

\makeatletter \@addtoreset{equation}{section} \makeatother

\def\text#1{\mbox{#1}}
\def\equ#1{(\ref{#1})}

\def\be#1{\begin{equation}\label{#1}}
\def\ee{\end{equation}}
\def\equ#1{(\ref{#1})}

\title{Topological mass generation to antisymmetric tensor matter field}
\author{L. Gonzaga Filho\inst{1,2}, M. S. Cunha\inst{2} and R. R. Landim\thanks{E-mail: \email{renan@fisica.ufc.br}}\inst{1}}
\institute{                                                              
  \inst{1} Departamento de
F\'{\i}sica, Universidade Federal do
Cear\'a, Caixa Postal 6030, 60455-760, Fortaleza, Cear\'a, Brazil\\                               \inst{2} N\'ucleo de F\'{\i}sica, Universidade Estadual do Cear\'a \\
Av. Paranjana, 1700, CEP 60740-000, Fortaleza, Cear\'a, Brazil           
}


\begin{document}
\maketitle

\begin{abstract}
We propose a mechanism to give mass to tensor matter field which preserve the $U(1)$ symmetry. We introduce a complex vector field that couples with the tensor in a topological term. We also analyze the influence of the kinetic terms of the complex vector in our mechanism.
\end{abstract}


Antisymmetric tensor fields have been object of study since a long time ago~\cite{ant}. Primarily they appear as a generalization of the vector gauge fields, i.e, have a gauge symmetry and their propagators possess zero modes.  This type of tensor fields, commonly called tensor gauge fields, permit us to construct topological invariants in $D$-dimensional manifolds \cite{schu,kalb} and have an important role in dualization \cite{spa1,spa2,hari1,wots}. Let us remark that such fields are the key to generate mass for the vector gauge fields through the topological mass mechanism \cite{lah,oda,landim1}.

A few years ago a new type of second rank antisymmetric tensor was introduced by Avdeev and Chizhov \cite{avdeev1}.  They constructed an action with spinors and vector gauge fields in which the tensor is a matter field rather than a gauge field and permit a quartic self-interacting invariant term. It exhibits
several interesting features among which we underline the asymptotically
free ultraviolet behavior of the abelian gauge interaction. Later Lemes {\it et al }\cite{lemes1} showed through BRST framework that this model is renormalizable into all perturbative orders. In a subsequent paper they realized that the tensor matter field is a real component of a complex tensor field that satisfies a complex self-dual condition \cite{lemes2}. As shown by the authors this condition makes the tensor matter field massless. Recently, Geyer and M\"{u}lsch \cite{geyer} constructed a supersymmetric topological $N_{T}\ge1$ model with the antisymmetric tensor matter fields. In that work they make an extension of the
Donaldson-Witten $N_{T}=1$ theory where the gauge field $A_{\mu }$ is now coupled to the anti-selfdual component of the tensor matter. They also obtained the super-BF $N_{T}=1$ model and the $N_{T}=2$ topological B-model.

The purpose of this paper is to study the possibility of giving mass to antisymmetric tensor matter field through a topological mechanism. This mechanism is similar to that one proposed by Allen, Bowick and Lahiri \cite{lah} making use of the topological $BF$ term. The main difference is that in  our mechanism we use a topological and a non-topological term with no gauge fields.

Let us begin introducing notations and conventions that
will be used through this paper. We use the metric $g_{\mu\nu}={\rm
diag}(+,-,-,-)$, and the completely antisymmetric tensor
$\varepsilon_{\mu\nu\rho\sigma}$ is defined using $\varepsilon_{0123}=1$. In
Minkowski space-time,
the antisymmetric tensor matter field $T_{\mu \nu }$ is a real component of
a complex second rank tensor $\varphi _{\mu \nu }$ which obeys the
complex self-dual condition~\cite{lemes2} namely,
\begin{equation}
i\widetilde{\varphi }_{\mu \nu }=\varphi _{\mu \nu },\quad \varphi
_{\mu \nu }=T_{\mu \nu }+i\widetilde{T}_{\mu \nu },  \label{eq1}
\end{equation}
where
\begin{equation}
\widetilde{\varphi }_{\mu \nu }=\frac 12\varepsilon _{\mu \nu \rho \sigma
}\varphi ^{\rho \sigma }.  \label{eq2}
\end{equation}
Let us emphasize here that a massive term with $U(1)$ symmetry for $\varphi_{\mu \nu }$ is forbidden due to the complex self-dual condition,
\begin{eqnarray}
\varphi_{\mu \nu }^\ast\varphi^{\mu \nu }=T_{\mu\nu}T^{\mu\nu}+\widetilde{T}_{\mu \nu }\widetilde{T}^{\mu \nu }=T_{\mu\nu}T^{\mu\nu}-T_{\rho\sigma}T^{\rho\sigma}=0,
\end{eqnarray}
where we used the fact that in the Minkowski space-time $\varepsilon_{\mu\nu\rho\sigma}\varepsilon^{\mu\nu\alpha\beta}=-\delta_{\rho\sigma}^{\alpha\beta}$, with $\delta_{\rho\sigma}^{\alpha\beta}=\delta^\alpha_\rho\delta^\beta_\sigma-\delta^\alpha_\sigma\delta^\beta_\rho$.

Since the complex self-dual condition makes the tensor $\varphi _{\mu \nu }$ massless, our aim is to construct a generating mass mechanism to the  tensor matter field that preserves the $U(1)$ symmetry  of the model.

We start with the following action
\begin{eqnarray}
S &=&\int d^{4}x\left[ -\left( D_{\mu }B_{\sigma }\right) ^{\ast}\left( D^{\mu
}B^{\sigma }\right) -\left( D_{\mu }\varphi ^{\mu \nu }\right) ^{\ast}
\left( D^{\sigma }\varphi _{\sigma \nu }\right) \right] \nonumber \\
&&+m\int d^{4}x\left[ B^{\sigma \ast }D^{\mu }\varphi _{\mu \sigma }+B^{\sigma
}\left( D^{\mu }\varphi _{\mu \sigma }\right) ^{\ast }\right], \label{acaoBT}
\end{eqnarray}
where $m$  is a mass parameter, $D_{\mu }$ represents the usual minimal
covariant derivative, $D_{\mu }=\partial _{\mu }-iA_{\mu },$ and $B_\mu$ is a complex vector field
\be{vectorB}
B_{\mu } =B_{\mu }^{\left( 1\right) }+iB_{\mu }^{\left( 2\right)}.
\ee
Clearly this action has an $U(1)$ local symmetry
\begin{eqnarray}
&&\delta B_{\mu}={i\alpha}B_\mu,\\
&&\delta \varphi_{\mu\nu}={i\alpha}\varphi_{\mu\nu},\\
&&\delta A_\mu=\partial_\mu\alpha.
\end{eqnarray}
Writing the action \equ{acaoBT} in terms of component fields, we have
\begin{eqnarray}
S &\approx&\int d^{4}x\left( -\partial _{\mu }B_{\sigma }^{\left( 1\right)
}\partial ^{\mu }B^{\left( 1\right)\sigma  }-\partial _{\mu }B_{\sigma
}^{\left( 2\right) }\partial ^{\mu }B^{\left( 2\right)\sigma }\right)
+J_{\sigma }^{\left( 1\right) }B^{\left( 1\right)\sigma }+J_{\sigma
}^{\left( 2\right) }B^{\left( 2\right)\sigma }  \nonumber \\
&&+\int d^{4}x\left( \frac{1}{2}\partial _{\sigma }T_{\mu \nu }\partial
^{\sigma }T^{\mu \nu }-2\partial ^{\mu }T_{\mu \nu }\partial _{\sigma
}T^{\,\,\sigma \nu }\right) +\frac{1}{2}J_{\mu \nu }T^{\,\,\mu \nu } \nonumber
\\
&&+2m\int d^{4}x\left( B^{\left( 1\right)\sigma }\partial ^{\mu }T_{\mu \sigma
}+B^{\left( 2\right) \sigma}\partial ^{\mu }\widetilde{T}_{\mu \sigma }\right)
,  \label{ação1}
\end{eqnarray}
where we are considering only the bi-linear terms and terms involving external sources. It is worth to mention that the last term is topological and it is linked to the mass parameter. The equations of motion
for the fields $T_{\mu \nu }$ and $B_{\mu }$ can be easily calculated by
minimal action principle, producing the equations below
\begin{eqnarray}
\partial ^{2}B^{\left( 1\right)\sigma}+m\partial _{\mu }T^{\,\mu
\sigma }+\frac12J^{\left( 1\right)\sigma} &=&0  \nonumber \\
\partial ^{2}B^{\left( 2\right)\sigma   }+m\partial _{\mu }\widetilde{T}%
^{\,\,\mu \sigma }+\frac12J^{\left( 2\right)\sigma } &=&0 \label{eqmovTB}
\end{eqnarray}
\vskip-0.4cm
\[
\partial^{2}T^{\rho \sigma }(x)-2\delta^{\rho\sigma}_{\mu\nu}\partial^\mu\partial_\lambda T^{\lambda\nu}(x)+m\delta^{\rho\sigma}_{\mu\nu}\partial^{\mu}B^{\nu\left(
1\right)}(x)+m\varepsilon ^{\rho \sigma \mu \lambda }\partial _{\mu }B_{\lambda
}^{\left( 2\right) }(x)-\frac12J^{\rho \sigma }(x)=0.
\]

In Feynman propagators calculation, we will use the generating functional
of the connected Green's functions $W$, with the following Legendre transformation
\begin{equation}
W[ T_{\mu \nu}, B_{\mu}^{(i)}] =\Gamma [ J_{\mu \nu},J_{\mu}^{(i)} ] -\frac12
\int dx'^4J_{\mu \nu}(x') T^{\mu \nu}(x')-\int dx'^4J_{\mu}^{(i)}(x') B^{\mu
(i)}(x').
\end{equation}
where the vacuum-to-vacuum transition amplitudes are defined by
\begin{eqnarray*}
\left\langle T_{\mu \nu}(x)\, T_{\rho \sigma}(y)\right\rangle=-i\frac{\delta^2 \Gamma }{\delta J_{\mu
\nu}(x)\delta J_{\rho \sigma}(y)} \Big|_{J=0}\\
\left\langle B_{\mu }^{(i)}(x)\, T_{\rho \sigma}(y)\right\rangle=-i\frac{\delta^2 \Gamma }{\delta
J_{\mu}(x)\delta J_{\rho \sigma}(y)}\Big|_{J=0}\,\, . \
\end{eqnarray*}
So, back to equations of motion (\ref{eqmovTB}), we obtain three expressions
to vacuum-to-vacuum transition amplitude for $B_{\mu }$ and $T_{\,\mu \nu }$
tensor matter fields, that represents the causal propagator in two
distinct point in space-time, viz, the two-point Green's function
\begin{equation}
\,\,\left\langle \partial ^{2}B_{\,}^{\left( 1\right)\sigma}\left( x\right)
T_{\alpha \beta }\left( y\right) \right\rangle +m\left\langle
\partial _{\mu }T^{\mu \sigma }\left( x\right) T_{\alpha \beta }\left( y\right) \right\rangle
=0 , \label{atvav1}
\end{equation}
\begin{equation}
\left\langle \partial ^{2}B^{\left( 2\right)\sigma}\left( x\right)
T_{\alpha \beta }\left( y\right) \right\rangle +m\left\langle \partial _{\mu
}\widetilde{T}^{\,\mu \sigma }\left( x\right)
T_{\alpha \beta }\left( y\right) \right\rangle =0  \label{atvav2}
\end{equation}
\[
2\left\langle \partial ^{2}T^{\,\rho \sigma }\left( x\right) T_{\alpha \beta
}\left( y\right) \right\rangle -4\delta^{\rho\sigma}_{\mu\nu}\left\langle \partial^\mu\partial_\lambda T^{\lambda\nu}(x) T_{\alpha \beta }\left(
y\right) \right\rangle +2m\delta^{\rho\sigma}_{\mu\nu}\left\langle \partial ^{\mu}B^{\left(
1\right)\nu}\left( x\right) T_{\alpha \beta }\left( y\right) \right\rangle
\]
\begin{equation}
+2m\,\varepsilon ^{\rho \sigma \mu \lambda }\left\langle \partial _{\mu
}B_{\,\lambda }^{\left( 2\right) }\left( x\right)
T_{\alpha \beta }\left( y\right) \right\rangle +i\,\delta _{\alpha
\beta }^{\,\rho \sigma }\delta ^{4}\left( x-y\right) =0 .
\label{atvav3}
\end{equation}

We must remember that our aim consists in determining the propagator for the
matter field $T_{\mu \nu }$. To do this we can work with the above expressions
eliminating  the fields $B_{\mu }^{\left( 1\right)}$ and $B_{\mu }^{\left( 2\right) }$. Doing so, we have
\begin{eqnarray}
\partial^{2}(\partial^{2}+m^2)\left\langle T^{\,\rho \sigma }\left( x\right)
T_{\alpha \beta }\left( y\right) \right\rangle
=i\,\delta^{\rho\sigma}_{\mu\nu}\delta^{\lambda\nu}_{\alpha\beta}\partial^\mu\partial_\lambda\delta^{4}\left( x-y\right) -\frac{i}{2}\partial
^{2}\delta _{\alpha \beta }^{\rho \sigma }\delta ^{4}\left( x-y\right).
\label{eqd2d2}
\end{eqnarray}
Using the Fourier representation for the Dirac delta function and Feynman propagator's,
\begin{eqnarray*}
\delta ^{4}\left( x-y\right) &=&\frac{1}{\left( 4\pi \right)
^{4}}\int
d^{4}k\,e^{ik\left( x-y\right) },  \nonumber \\
\left\langle T^{\rho \sigma }\left( x\right) T_{\alpha \beta
}\left( y\right) \right\rangle &=&\frac{1}{\left( 4\pi \right)
^{4}}\int d^{4}k\,G_{\alpha \beta }^{\rho \sigma }\left( k\right)
\,e^{ik\left( x-y\right) },
\end{eqnarray*}
we obtain the following expression for Green's function
\begin{eqnarray}
G_{\alpha \beta }^{\rho \sigma }\left( k\right) &=&\frac{i}{\left(
k^{2}-m^{2}\right) }\Pi _{\alpha \beta }^{\rho \sigma }\left( k\right),
\end{eqnarray}
with
\[
\Pi _{\alpha \beta }^{\rho \sigma }\left( k\right)
=\frac{1}{2}\delta _{\alpha \beta }^{\rho\sigma}-\frac{1}{k^{2}}\delta^{\rho\sigma}_{\mu\nu}\delta^{\lambda\nu}_{\alpha\beta}k^\mu k_\lambda.
\]

It is worth to observe that the causal propagator for the antisymmetric tensor
matter field has a massive pole due to the topological mass, which comes from
the coupling between  $T_{\mu \nu }$ and the vector field $B_{\mu }$. We can
compare this result with the one obtained in Avdeev and Chizhov work. There, we
can see only one massless pole propagator for the field $T_{\mu \nu}$ \cite{avdeev1}.

Now let us construct a more general  bilinear action in the  fields $B_{\mu }$
and $T_{\mu \nu }$. To do this, let us introduce a new kinetic term, namely, $a \left( D_{\mu }B^{\mu }\right) ^{\ast }\left( D_{\sigma }B^{\sigma
}\right) $, with $a$ representing a real constant to be determined, and a mass term for $B_\mu$ fields.  We
will see how these additional terms can affect causal propagator for the matter field $T_{\mu \nu }$. Thus, we have a new action $S$,
\begin{eqnarray}
S &=&\int d^{4}x\left[ -\left( D_{\mu }B_{\sigma }\right) ^{\ast
}\left( D^{\mu }B^{\sigma }\right) +a\left( D_{\mu }B^{\mu
}\right) ^{\ast }\left( D_{\sigma }B^{\sigma }\right) -\left(
D_{\mu }\varphi ^{\mu \nu }\right) ^{\ast }\left( D^{\sigma }\varphi
_{\sigma \nu }\right) \right]
\nonumber \\
&&+\int d^{4}x\left[m\left(B^{\nu \ast }D^{\mu }\varphi _{\mu \nu }+B^{\nu}\left(
D^{\mu }\varphi _{\mu \nu }\right) ^{\ast }\right)+\mu^2B_\sigma^\ast B^\sigma\right]. \label{actionadcnal}
\end{eqnarray}
Writing it in components again and considering only the bi-linear and source terms, we obtain
the following equations of motion
\begin{eqnarray}
\partial ^{2}B^{\left( 1\right)\sigma  }-a\partial _{\mu }\partial
^{\sigma }B^{\left( 1\right)\mu }+\mu^2 B^{(1)\sigma}+m\partial _{\mu }T^{\mu \sigma } + \frac 12
J^{\left( 1\right)\sigma} &=&0 \nonumber,\\  
\partial ^{2}B_{\sigma }^{\left( 2\right)}-a
\partial _{\mu
}\partial ^{\sigma }B^{\left( 2\right) \mu }+\mu^2B^{\left( 2\right) \mu }+m\partial _{\mu }\widetilde{T}%
^{\mu \sigma }+\frac12 J^{\left( 2\right) \sigma }\label{eqmov2} &=&0,
\end{eqnarray}
\vskip -.5cm
\[
\partial^{2}T^{\rho \sigma }(x)-2\delta^{\rho\sigma}_{\mu\nu}\partial^\mu\partial_\lambda T^{\lambda\nu}(x)+m\delta^{\rho\sigma}_{\mu\nu}\partial^{\mu}B^{\nu\left(
1\right)}(x)+m\varepsilon ^{\rho \sigma \mu \lambda }\partial _{\mu }B_{\lambda
}^{\left( 2\right) }(x)-\frac12J^{\rho \sigma }(x)=0.
\]
Let us observe that the first and second equations above have acquired two new
terms compared with equations (\ref{eqmovTB}). However, the last equation
remains the same. From the equations above we obtain the following
expressions for the vacuum-to-vacuum transition amplitudes:

\begin{eqnarray}
&&\left<T_{\mu\nu}(x)T^{\rho\sigma}(y)\right>=\frac{1}{(2\pi)^4}\int G_{\mu\nu}^{\rho\sigma}(k)e^{ik(x-y)}d^4k,\nonumber\\
&&\left<B^{(1)}_\mu(x)B^{(1)}\nu(y)\right>=\left<B^{(2)}_\mu(x)B^{(2)\nu}(y)\right>=\frac{1}{(2\pi)^4}\int G_\mu^\nu(k)e^{ik(x-y)}d^4k,\nonumber\\
&&\left<B^{(1)}_\mu(x)B^{(2)}\nu(y)\right>=\frac{1}{(2\pi)^4}\int G_\mu^{(12)\nu}(k)e^{ik(x-y)}d^4k,\\
&&\left<B^{(1)}_\mu(x)T^{\rho\sigma}(y)\right>=\frac{1}{(2\pi)^4}\int \Gamma^{(1)\rho\sigma}_\mu(k)e^{ik(x-y)}d^4k,\nonumber\\
&&\left<B^{(2)}_\mu(x)T^{\rho\sigma}(y)\right>=\frac{1}{(2\pi)^4}\int \Gamma^{(2)\rho\sigma}_\mu(k)e^{ik(x-y)}d^4k,\nonumber
\end{eqnarray}
with
\begin{eqnarray}
&&G_{\mu\nu}^{\rho\sigma}(k)=\frac{i(k^2-\mu^2)}{k^2(k^2-\mu^2-m^2)}\Pi^{\rho\sigma}_{\mu\nu}(k),\nonumber\\
&&G_\mu^\nu(k)=\frac{im^2k_\sigma k^\rho\delta^{\sigma\nu}_{\rho\mu}}{2k^2(k^2-\mu^2-m^2)(k^2-\mu^2)}+\frac{iak_\mu k^\nu}{2((1-a)k^2-\mu^2)}-\frac{i\delta_\mu^\nu}{2(k^2-\mu^2)}\nonumber\\
&&G_\mu^{(12)\nu}=\frac{im^2k_\sigma k^\rho\delta^{\sigma\nu}_{\rho\mu}}{2k^2(k^2-\mu^2-m^2)(k^2-\mu^2)}\\
&&\Gamma^{(1)\rho\sigma}_\mu(k)=\frac{-mk^\nu\delta^{\rho\sigma}_{\mu\nu}}{2k^2(k^2-m^2-\mu^2)}\nonumber,\\
&&\Gamma^{(2)\rho\sigma}_\mu(k)=\frac{mk^\nu\epsilon^{\rho\sigma}_{\mu\nu}}{2k^2(k^2-m^2-\mu^2)}\nonumber.
\end{eqnarray}

We obtain an interesting result:  the presence of the term $a \left( D_{\mu }B^{\mu }\right) ^{\ast }\left( D_{\sigma }B^{\sigma
}\right) $,   affects only  the amplitude $G_\mu^\nu(k)$. The tensor matter field propagator's has a massive pole which does not depend on the parameter $a$.

In conclusion, we propose a mechanism to generate mass to antisymmetric tensor matter field which preserves $U(1)$ symmetry. In this mechanism a topological term is introduced via a complex vector field and the complex self-dual condition. Whereas this condition makes the tensor massless, the coupling with the vector field gives a massive pole to the tensor propagator.

\vskip1cm
  \noindent
  We wish to thank Marcio A. M. Gomes for reading the manuscript.
 Conselho Nacional de Desenvolvimento Cient\'\i fico e tecnol\'ogico-CNPq is gratefully
acknowledged for financial support.

\end{document}